\documentclass[preprint]{ptephy_v1}

\preprintnumber{XXXX-XXXX} 

\usepackage{amsmath,cases}
\usepackage{lastpage}
\usepackage{graphicx}
\usepackage{epstopdf}
\usepackage{mathtools}
\usepackage{booktabs}
\usepackage{braket}
\usepackage{comment}

\usepackage{scalerel}
\usepackage{tikz}
\usepackage[hang,small,bf]{caption}
\usepackage[subrefformat=parens]{subcaption}
\usepackage{multirow}
\usepackage{lineno}
\usepackage{hyperref}
\usepackage{siunitx}

\newcommand{\Var}{\mathrm{Var}}

\newcommand{\Cov}{\mathrm{Cov}}
\newcommand{\Nside}{$\mathrm{N_{side}}$}
\newcommand{\ellmin}{\ell_\mathrm{min}}
\newcommand{\ellmax}{\ell_\mathrm{max}}
\newcommand{\bell}{\mathbf{l}}
\newcommand{\btheta}{\hat{\mathbf{\theta}}}

\nolinenumbers

\begin{document}

\title{Determination of miscalibrated polarization angles from observed CMB and foreground $EB$ power spectra: Application to partial-sky observation}

\author[1,*]{Yuto Minami}
\affil[1]{High Energy Accelerator Research Organization, 1-1 Oho, Tsukuba, Ibaraki 305-0801, Japan \email{yminami@post.kek.jp}}

\begin{abstract}%
We study a strategy to determine miscalibrated polarization angles of cosmic microwave background (CMB) experiments using the observed $EB$ polarization power spectra of CMB and Galactic foreground emission.
We apply the methodology of Ref.~\cite{Minami:2019ruj} developed for full-sky observations to ground-based experiments such as Simons Observatory.
We take into account the $E$-to-$B$ leakage and $\ell$-to-$\ell$ covariance due to partial sky coverage using the public code NaMaster.
We show that our method yields an unbiased estimate of miscalibrated angles.
Our method also enables simultaneous determination of miscalibrated angles and the intrinsic $EB$ power spectrum of polarized dust emission
when the latter is proportional to $\sqrt{C_\ell^{EE}C_\ell^{BB}}$ and $C_\ell^{BB}$ is proportional to $C_\ell^{EE}$.
\end{abstract}

\subjectindex{xxxx, xxx}

\maketitle

\section{Introduction}\label{sec:Introduction}
$B$-mode polarization of the cosmic microwave background (CMB) is 
a probe of primordial gravitational waves generated during inflation~\cite{Kamionkowski:2015yta}.
To detect this, we need to remove spurious $B$-mode signals such as those from gravitational lensing~\cite{Zaldarriaga:1998ar}, patchy reionization~\cite{Hu:1999vq,Dvorkin:2009ah}, $E$-to-$B$ leakage due to partial sky coverage~\cite{Bunn:2002df,Smith:2006vq},
and miscalibration of polarization angles of detectors~\cite{Komatsu:2010fb}. 
In this paper, we focus on estimation of the miscalibrated polarization angles from CMB experiments.

The miscalibrated polarization angles, $\alpha$, yield not only a spurious power spectrum of the $B$-mode polarization,
but also a cross power spectrum of $E$- and $B$-mode polarization~\cite{Komatsu:2010fb}.
The traditional method to estimate $\alpha$ is to relate the observed $EB$ power spectrum to the \textit{theoretical} model of the difference between the $E$- and $B$-mode power spectra of CMB~\cite{Keating:2012ge}.
This approach is thus not applicable in foreground-dominated frequency bands.
In Ref.~\cite{Minami:2019ruj}, we solved this problem by relating the observed $EB$ power spectrum to the \textit{observed} difference between the $E$- and $B$-mode power spectra in the sky.
However, the analysis given there was limited to full-sky observations by a satellite experiment such as LiteBIRD~\cite{Hazumi2019}.
In this paper, we apply the methodology developed in Ref.~\cite{Minami:2019ruj} to ground-based observations with a partial sky coverage.
While we use Simons Observatory~\cite{Ade:2018sbj} as an example,
our method can be applied to any other experiments including Simons Array~\cite{Ben:2018}, BICEP Array~\cite{Hui:2018cvg}, and CMB-S4~\cite{abazajian2019cmbs4}.

Partial sky breaks orthogonality of spin-2 spherical harmonics,
resulting in the leakage of $E$-mode to $B$-mode and the correlation between different multipoles.
In this paper we take these effects into account by using the public code NaMaster~\cite{Alonso:2018jzx}.

The rest of the paper is organized as follows.
In Sect.~\ref{sec:Methodology} we review the methodology of Ref.~\cite{Minami:2019ruj} and describe the extension to partial-sky data. 
In Sect.~\ref{sec:SkySim} we describe the way to generate sky simulations for validation of our method. 
In Sect.~\ref{sec:Results} we validate our method using sky simulations,
and present the main results.
In Sect.~\ref{sec:ForegroundEB} we extend our method to determine the miscalibration angle and the intrinsic $EB$ correlation of thermal dust emission simultaneously.
We conclude in Sect.~\ref{sec:Conclusion}.
\section{Methodology}\label{sec:Methodology}

\subsection{Review of the methodology}
We review the methodology of Ref.~\cite{Minami:2019ruj} developed for full-sky observations.
In this section,
all spherical harmonics coefficients and power spectra are calculated from full-sky data without a mask.

In Ref.~\cite{Minami:2019ruj},
we took into account both the miscalibration angle, $\alpha$,
and the ``cosmic birefringence''~\cite{Carroll:1998zi,Lue:1998mq,Feng:2004mq,Feng:2006dp,Liu:2006uh} angle, $\beta$, which only rotates CMB polarization.
When we consider noise (``N''), foreground (``fg''), and CMB (``CMB'') components,
spherical harmonics coefficients of the observed ('o') $E$- and $B$- mode polarization
are related to the intrinsic ones by
\begin{align}
E_{\ell,m}^\mathrm{o} &= 
E_{\ell,m}^\mathrm{fg}\cos(2\alpha)-  B_{\ell,m}^\mathrm{fg}\sin(2\alpha)
+E_{\ell,m}^\mathrm{CMB}\cos(2\alpha+2\beta)-  B_{\ell,m}^\mathrm{CMB}\sin(2\alpha+2\beta)
+E_{\ell,m}^\mathrm{N},
\\
B_{\ell,m}^\mathrm{o} &= 
E_{\ell,m}^\mathrm{fg}\sin(2\alpha) + B_{\ell,m}^\mathrm{fg}\cos(2\alpha)
+E_{\ell,m}^\mathrm{CMB}\sin(2\alpha+2\beta) + B_{\ell,m}^\mathrm{CMB}\cos(2\alpha+2\beta)
+B_{\ell,m}^\mathrm{N}.
\end{align} 
In this paper,
we use the notations that all spherical harmonics coefficients and power spectra have been multiplied by the appropriate beam smoothing function. 

When we define the power spectra with ensemble average as $\langle C_\ell^{XY} \rangle=(2\ell+1)^{-1}\sum_{m=-\ell}^\ell \langle X_{\ell,m}Y_{\ell,m}^* \rangle$, 
we obtain
\begin{equation}\label{eq:GeneralRotationFitting}
\begin{split}
\langle C_\ell^{EB,\mathrm{o}}\rangle =&
\frac{\tan(4\alpha ) }{2}
\left(\langle C_\ell^{EE,\mathrm{o}}\rangle-\langle C_\ell^{BB,\mathrm{o}}\rangle\right)
+
\frac{\sin(4\beta)}{2\cos(4\alpha)}\left(
\langle C_\ell^{EE,\mathrm{CMB}}\rangle - \langle C_\ell^{BB,\mathrm{CMB}}\rangle
\right)\\
&+\frac1{\cos(4\alpha)}\langle C_\ell^{EB,\mathrm{fg}}\rangle
+\frac{\cos(4\beta)}{\cos(4\alpha)}\langle C_\ell^{EB,\mathrm{CMB}}\rangle\,.
\end{split}
\end{equation}

The current data show no evidence for non-zero $EB$ correlation from the foreground emission~\cite{planckdust:2016,planckdust:2018} and from the CMB~\cite{Aghanim:2016fhp}.
Ignoring $\langle C_\ell^{EB,\mathrm{fg}}\rangle$ and $\langle C_\ell^{EB,\mathrm{CMB}}\rangle$,
we can determine $\alpha$ and $\beta$ simultaneously. 

It is however possible that future experiments may find a non-zero $EB$ correlation. 
We revisit the term $\langle C_\ell^{EB,\mathrm{fg}}\rangle$ in Sect.~\ref{sec:ForegroundEB} for the case that we do not ignore $\langle C_\ell^{EB,\mathrm{fg}}\rangle$.

To determine $\alpha$ and $\beta$ simultaneously, 
we used a log-likelihood function given by
\begin{equation}\label{eq:LikelihoodGeneral}
-2\ln \mathcal{L}
= \sum_{\ell=2}^{\ell_\mathrm{max}}
\frac{
	\left[
	C_\ell^{EB,\mathrm{o} } 
	- \frac{\tan(4\alpha) }{ 2 }  \left( C_\ell^{EE,\mathrm{o} } - C_\ell^{ BB,\mathrm{o} } \right) 	
	-  \frac{ \sin(4\beta) }{ 2\cos( 4\alpha ) }
	\left(
	C_{\ell}^{EE,\mathrm{CMB,th} }b_\ell^2 - C_{\ell}^{BB,\mathrm{CMB,th} }b_\ell^2 
	\right)	
	\right]^2
}{
	\Var \left( 
	C_\ell^{ EB, \mathrm{o} } - \frac{\tan(4\alpha) }{ 2 }  \left( C_\ell^{EE,\mathrm{o} } - C_\ell^{ BB,\mathrm{o} } \right) 
	\right)
},
\end{equation}
where $C_{\ell}^{EE,\mathrm{CMB,th} }b_\ell^2$ and $C_{\ell}^{BB,\mathrm{CMB,th} }b_\ell^2$ are the best-fitting $\Lambda$CDM theoretical power spectra multiplied by the beam transfer functions, $b_\ell^2$.
As for the variance in the denominator,
we used an approximated variance given by~\cite{Minami:2019ruj}
\begin{equation}
\begin{split}
&\Var \left( C_\ell^{EB, \mathrm{o}}   - ( C_\ell^{EE, \mathrm{o}} - C_\ell^{BB, \mathrm{o}})\tan(4\alpha)/2 \right) \\
&\approx
\frac{1}{2\ell+1}C_\ell^{EE,\mathrm{o}}C_\ell^{BB,\mathrm{o}}
+\frac{\tan^2(4\alpha)}{4}\frac{2}{2\ell+1}\left[
(C_\ell^{EE,\mathrm{o}})^2
+
(C_\ell^{BB,\mathrm{o}})^2
\right]
\\
&\quad
-\tan(4\alpha)\frac{2}{2\ell+1}C_\ell^{EB,\mathrm{o}}
\left( C_\ell^{EE,\mathrm{o}} - C_\ell^{BB,\mathrm{o}} 
\right).
\end{split}
\end{equation}
To determine $\alpha$ and $\beta$, we minimized Eq.~\ref{eq:LikelihoodGeneral} with respect to $\alpha$ and $\beta$,
given $C_\ell^{ EB, \mathrm{o}}$, $\left(C^{EE,\mathrm{o}}-C^{BB,\mathrm{o}}\right)$, $C_\ell^{EE,\mathrm{CMB,th}}$, and $C_\ell^{BB,\mathrm{CMB,th}}$.

\subsection{Extension to partial-sky data}
In this paper, we shall extend the methodology of Ref.~\cite{Minami:2019ruj} to partial-sky data.
We focus on determination of $\alpha$ and ignore the cosmic birefringence angle, $\beta$, in Eq.~\ref{eq:LikelihoodGeneral}.
We obtain
\begin{equation}\label{eq:LikelihoodAlphaDiagonal}
-2\ln \mathcal{L}
= \sum_{\ell= \ell_\mathrm{min}}^{\ell_\mathrm{max}}
\frac{
	\left[
	C_\ell^{EB,\mathrm{o} } 
	- \frac{\tan(4\alpha) }{ 2 }  \left( C_\ell^{EE,\mathrm{o} } - C_\ell^{ BB,\mathrm{o} } \right) 	
	\right]^2
}{
	\Var \left( 
	C_\ell^{ EB, \mathrm{o} } - \frac{\tan(4\alpha) }{ 2 }  \left( C_\ell^{EE,\mathrm{o} } - C_\ell^{ BB,\mathrm{o} } \right) 
	\right)
}.
\end{equation}
However, Eq.~(\ref{eq:LikelihoodAlphaDiagonal}) ignores the $\ell$-to-$\ell$ bin covariance due to partial sky coverage.
We thus write
\begin{equation}\label{eq:LikelihoodAlphaNonDiagonal}
-2\ln \mathcal{L}
= \vec{C}^T(\alpha) \mathbf{C}(\alpha)^{-1} \vec{C}(\alpha) + \ln\left|\mathbf{C}(\alpha)\right|
,
\end{equation}
where $\vec{C}(\alpha)$ is a one-dimensional array of $C_\ell^{EB,\mathrm{o} } 
- \frac{\tan(4\alpha) }{ 2 }  \left( C_\ell^{EE,\mathrm{o} } - C_\ell^{ BB,\mathrm{o} } \right) $
with $\ell$ from $\ell_\mathrm{min}$ to $\ell_\mathrm{max}$,
and $\mathbf{C}(\alpha)$ is the corresponding covariance matrix.
The explicit form of $\mathbf{C}(\alpha)$ is
\begin{equation}\label{eq:Covariance}
\begin{split}
&\mathbf{C}(\alpha)_{\ell,\ell^{'}} = 
\Cov \left(C_\ell^{EB, \mathrm{o}}, C_{\ell^{'}}^{EB, \mathrm{o}}\right) 
\\&+
\frac{\tan^2(4\alpha)}{4} \left[
\Cov \left( C_\ell^{EE, \mathrm{o}}, C_{\ell^{'}}^{EE, \mathrm{o}} \right) + \Cov \left( C_\ell^{BB, \mathrm{o}} , C_{\ell^{'}}^{BB, \mathrm{o}}\right)
-\Cov \left(  C_\ell^{EE, \mathrm{o}} , C_{\ell^{'}}^{BB, \mathrm{o}} \right)
- \Cov \left(  C_\ell^{BB, \mathrm{o}} , C_{\ell^{'}}^{EE, \mathrm{o}} \right)
\right]
\\
&
-\frac{\tan(4\alpha)}{2}
\left[
\Cov\left( C_\ell^{EB, \mathrm{o}} , C_{\ell^{'}}^{EE, \mathrm{o}} - C_{\ell^{'}}^{BB, \mathrm{o}} \right)
+ 
\Cov\left( C_{\ell}^{EE, \mathrm{o}} - C_{\ell}^{BB, \mathrm{o}},  C_{\ell^{'}}^{EB, \mathrm{o}}  \right)
\right].
\end{split}
\end{equation}
The size of $\mathbf{C}(\alpha)$ is $(\ellmax-\ellmin+1)$-by-$(\ellmax-\ellmin+1)$.
To determine $\alpha$, 
we minimize Eq.~(\ref{eq:LikelihoodAlphaNonDiagonal}) with respect to $\alpha$,
given $C_\ell^{ EB, \mathrm{o}}$ and $\left(C_\ell^{EE,\mathrm{o}}-C_\ell^{BB,\mathrm{o}}\right)$.

\subsection{Estimation of unbiased power spectra and their covariance matrices}\label{sec:NaMaster}
Throughout this paper,
we use a public code ``NaMaster''~\cite{Alonso:2018jzx} to estimate unbiased $C_\ell$ from partial-sky data.
Since we can describe the $E$-to-$B$ leakage and $\ell$-to-$\ell$ covariance due to a partial sky coverage by the spherical harmonics transform of the sky mask,
we can correct them and estimate the unbiased $C_\ell$ from partial-sky data~\cite{Hivon:2002,Elsner:2016,Alonso:2018jzx}.

We also use NaMaster to estimate the covariance matrices of all pairs of $C_\ell^{EE, \mathrm{o}}$, $C_\ell^{BB, \mathrm{o}}$, and $C_\ell^{EB, \mathrm{o}}$.
NaMaster estimates the covariance matrices under certain approximations~\cite{Efstathiou:2004,Brown:2004jn}.
We validate the estimated covariance matrices against Monte-Carlo (MC) simulations
in Appendix~\ref{sec:ValidCov}.

\section{Sky simulations}\label{sec:SkySim}
To validate our methodology,
we use the ``PySM'' package~\cite{Thorne:2016ifb} to produce realistic simulations of the microwave sky,
with an experimental specification similar to the Large Aperture Telescope (LAT) of Simons Observatory (SO)~\cite{Ade:2018sbj} (see  Table~\ref{tab:SOParameter}).
We include two polarized Galactic foreground emission models:
``s1'' for synchrotron model and ``d1'' for dust emission model,
as described in Ref.~\cite{Thorne:2016ifb}.
The noise is assumed to be white with standard deviation given by 
$\sigma_\mathrm{N} = (\pi/10800)(w_{\rm p}^{-1/2}/\mu{\rm K~arcmin})~\mu \mathrm{K~str^{-1/2}} $~\cite{Katayama:2011eh}
with $w_{\rm p}^{-1/2}$ given in the ``Polarization Sensitivity'' row of Table~\ref{tab:SOParameter}.
A CMB map is generated from the power spectra calculated by CAMB~\cite{Lewis:2000}
using the latest Planck 2018 cosmological parameters for ``TT,TE,EE$+$lowE$+$lensing''~\cite{Aghanim:2018eyx}:
$\Omega_bh^2=0.02237$, $\Omega_ch^2=0.1200$,
$h=0.6736$, $\tau=0.0544$, $A_s=2.100\times 10^{-9}$, and $n_s=0.9649$.
The CMB map includes the lensed $B$-mode but does not include the primordial $B$-mode, i.e., the tensor-to-scalar ratio, $r$, is zero.

We choose a frequency band of $280\,\si{\GHz}$ for this study
because it is the highest frequency of the SO LAT
and is most affected by the polarized dust emission.
The generated map is masked by LR42 galactic mask of Ref.~\cite{planckdust:2018},
because its sky coverage, $f_\mathrm{sky} = 0.42$, is similar to that of the SO LAT, $f_\mathrm{sky} = 0.4$~\cite{Ade:2018sbj}.
The edges of the mask are apodized with a $5\,\deg$ FWHM Gaussian to reduce the mask induced $E$-to-$B$ leakage.

In this paper we set the input miscalibration angle $\alpha$ ($\alpha_\mathrm{in}$) to $0.33\deg$,
which corresponds to the calibration uncertainty of the Crab nebula (Tau A)
~\cite{Aumont:2018epb}.

In the following sections,
we will show results both with a single realization and with MC simulations.
With a single realization,
we will test whether we can recover the input angles within the estimated uncertainties.
With MC simulations,
we will test whether the estimator is unbiased and whether the estimated uncertainty is valid.

For the MC simulations,
we have generated many different realizations assuming Gaussian fluctuations with zero mean.
The Gaussian foreground realizations have been generated from the power spectra of the PySM map,
while the CMB and noise realizations have been generated with the same way described above.
We have used the \texttt{synfast} function of HEALPix to generate Gaussian realizations from the input power spectra.

\begin{table}
	\centering
	\caption{
		Experimental parameters similar to the Simons Observatory Large Aperture Telescope~\cite{Ade:2018sbj}. 
	}
	\label{tab:SOParameter}
	\begin{tabular}{c c}
		\toprule
		Parameters & values\\
		\midrule
		Frequency band & $280\,\si{\GHz}$\\
		\Nside of map &  2048\\
		$(\ellmin, \ellmax)$ & $(200, 5000)$ \\
		Polarization sensitivity ($w_\mathrm{p}^{-1/2}$ ) & 54.0 $\mu\mathrm{K}$-$\mathrm{arcmin}$\\
		Beam width (at FWHM) &  $0.9$\,arcmin\\
		$f_\mathrm{sky}$ & $0.42$ \\
		\bottomrule
	\end{tabular}
\end{table}
\section{Results}\label{sec:Results}
First, we report the results of $\alpha$ determination 
from a simpler (but sub-optimal) log-likelihood given by Eq.~(\ref{eq:LikelihoodAlphaDiagonal}).
As the off-diagonal elements of the covariance matrix 
have negative values (see Appendix~\ref{sec:ValidNonDiagonal}),
neglecting off-diagonal elements gives a conservative uncertainty in $\alpha$.
We show recovered $\alpha$ and their $1\sigma$ uncertainties derived from a single realization
and MC simulations with 100 realizations in the left two columns of
Table~\ref{tab:alpha}.

For the results with a single realization,
we show the best fitting values which make the log-likelihood minimum
and their estimated uncertainty which increase the log-likelihood,
$-2\ln\mathcal{L}$, by one from the minimum.
For the results with MC simulations,
we show the means and the standard deviations of the best fitting values of all the MC samples.

We find that the sub-optimal method recovers correctly $\alpha_\mathrm{in}$,
whereas it overestimates uncertainty.
Thus, we can use our simple (but sub-optimal) method as a conservative method.

As a reference, 
we also show the results from the traditional method~\cite{Keating:2012ge}
in the right two columns of Table~\ref{tab:alpha} 
using the log-likelihood function of 
\begin{equation}\label{eq:LikelihoodTraditional}
-2\ln \mathcal{L}
= \sum_{\ell=\ellmin}^{\ellmax}
\frac{
	\left[
	C_\ell^{EB,\mathrm{o} } 
	- \frac{\sin(4\alpha) }{ 2 }  \left( C_\ell^{EE,\mathrm{CMB,th} }b_\ell^2 - C_\ell^{ BB,\mathrm{CMB,th} }b_\ell^2 \right) 	
	\right]^2
}{
	\Var \left(  C_\ell^{ EB, \mathrm{o} }\right)
}.
\end{equation}
Here the traditional method relates the observed $EB$ power spectrum to the theoretical model of the difference between the $E$- and $B$-mode power spectra of CMB.
As for $\Var \left(  C_\ell^{ EB, \mathrm{o} }\right)$, 
we use the diagonal elements of the covariance matrix, $\Cov \left(C_\ell^{EB, \mathrm{o}}, C_{\ell^{'}}^{EB, \mathrm{o}}\right)$, estimated by NaMaster. 
Since this log-likelihood also ignores the $\ell$-to-$\ell$ bin covariance,
it is also a sub-optimal method.

There are two notable points:
(1) With a single realization, 
the estimated uncertainty from our sub-optimal method is smaller than 
that from the traditional sub-optimal method
because our sub-optimal method uses correlation between the observed power spectra; 
(2) The estimator from our sub-optimal method is unbiased while that from the sub-optimal traditional method is biased.
This is because our method automatically includes foreground components,
while the traditional method ignores them.

\begin{table}
	\centering
	\caption{
		Recovered $\alpha$ and their $1\sigma$ uncertainties against $\alpha_\mathrm{in}=0.33\,\deg$ with two methods:
		our simpler (but sub-optimal) method (\ref{eq:LikelihoodAlphaDiagonal}) and the traditional sub-optimal method (\ref{eq:LikelihoodTraditional}).
		In both methods, 
		we show results of both a single realization and MC simulations.
	}
	\label{tab:alpha}
	\begin{tabular}{ccc c}
		\toprule
		\multicolumn{2}{c}{$\alpha$ (deg) from Eq.~(\ref{eq:LikelihoodAlphaDiagonal})} &\multicolumn{2}{c}{	$\alpha$ (deg) from Eq.~(\ref{eq:LikelihoodTraditional})}\\
		\cmidrule(lr){1-2} \cmidrule(lr){3-4}	
		Single realization & MC simulations & Single realization & MC simulations\\
		\cmidrule(lr){1-2} \cmidrule(lr){3-4}	
		$0.33\pm 0.17$ & $0.34\pm 0.13$  &  $0.41 \pm 0.29$ & $0.54 \pm 0.13$\\
		\bottomrule
	\end{tabular}
\end{table}

\begin{table}
	\centering
	\caption{
		Recovered $\alpha$ and its $1\sigma$ uncertainty against input $\alpha=0.33\,\deg$ from our optimal method (\ref{eq:LikelihoodAlphaNonDiagonal}).
			We show results of both a single realization and MC simulations.
	}
	\label{tab:alphaCov}
	\begin{tabular}{cc}
		\toprule
		\multicolumn{2}{c}{	$\alpha$ (deg) from Eq.~(\ref{eq:LikelihoodAlphaNonDiagonal}) } \\
		\midrule
		Single realization & MC simulations\\
		\midrule
		$0.302\pm 0.085$ & $0.330 \pm 0.086$ \\
		\bottomrule
	\end{tabular}
\end{table}

Next,
we report the results from the optimal log-likelihood as given in Eq.~(\ref{eq:LikelihoodAlphaNonDiagonal}).
We set the bin size to $\Delta\ell = 8$, which leads to $600$-bins from $\ellmin=200$ to $\ellmax=5000$.
We show the recovered values of $\alpha$ and their $1\sigma$ uncertainty in 
Table~\ref{tab:alphaCov}.
In a similar way with the sub-optimal methods,
we derive results of both a single realization and MC simulations with 1000 realizations.

We find that our optimal method recovers correctly $\alpha_\mathrm{in}$ with a single realization and
find that the estimator is unbiased.
We also find that the 
the covariance matrix in Eq.~(\ref{eq:LikelihoodAlphaNonDiagonal}) is valid
because the uncertainty estimated by a single realization
is consistent with the uncertainty derived from MC simulations.

The uncertainty on $\alpha$ is smaller than the uncertainty determined only with the diagonal elements of covariance matrix (\ref{eq:LikelihoodAlphaDiagonal}).
This is because the off-diagonal covariance-matrix elements have negative values as described in Appendix~\ref{sec:ValidNonDiagonal}.

\section{Intrinsic $EB$ correlation of Galactic foreground emission}
\label{sec:ForegroundEB}
In this section,
we show that we can determine the miscalibration angle, $\alpha$, and the intrinsic $EB$ power spectrum of polarized dust emission (if any) simultaneously.
We suggested this idea in Ref.~\cite{Minami:2019ruj};
The key is to put thermal dust $EB$ correlation parameters into a new rotation angle, $\gamma$.

In a simple model~\cite{Abitbol:2015epq},
we can relate $\langle C_\ell^{EB,fg}\rangle$ to $\langle C_\ell^{EE,fg}\rangle$ and $\langle C_\ell^{BB,fg}\rangle$ by
\begin{equation}\label{eq:EBtoEEBB}
\langle C_\ell^{EB,fg} \rangle = f_c \sqrt{\langle C_\ell^{EE,fg} \rangle \langle C_\ell^{BB,fg}\rangle},
\end{equation}
where $f_c$ is a correlation coefficient.
Since the dust foreground $EE$ and $BB$ power spectra 
have a similar shape in $\ell$~\cite{planckdust:2018},
we may write the $BB$ spectrum as 
\begin{equation}\label{eq:BBpropEE}
\langle C_\ell^{BB,fg}\rangle = \xi \langle C_\ell^{EE,fg}\rangle, 
\end{equation}
where $\xi$ is, e.g., equal to $\langle A_{BB}/A_{EE}\rangle\approx 0.5$~\cite{planckdust:2018}.
When $f_c$ satisfies $0 \leq 2 f_c \sqrt{\xi}/(1-\xi) \leq 1$,
we can parametrize the foreground $EB$ correlation with a rotation angle, $\gamma$, as $\sin(4\gamma)/2  = f_c \sqrt{\xi}/(1-\xi) $. 
Then we have~\cite{Minami:2019ruj}
\begin{equation}
\langle C_\ell^{EB,\mathrm{fg}} \rangle = \frac{\sin(4\gamma)}{2} \left(\langle C_\ell^{EE,\mathrm{fg}} \rangle-\langle C_\ell^{BB,\mathrm{fg}} \rangle\right).
\end{equation}
Thus we can put thermal dust $EB$ correlation parameters into a rotation angle, $\gamma$.

Since dust foreground is rotated by $\alpha + \gamma$ and CMB is rotated by $\alpha$,
we can use Eq.~(\ref{eq:LikelihoodGeneral}) to determine $\alpha$ and $\gamma$ simultaneously,
if we replace $\alpha \rightarrow \alpha+\gamma$ and $\beta \rightarrow -\gamma $.
For partial-sky data, we have
\begin{equation}\label{eq:LikelihoodAlphaGammaNonDiagonal}
-2\ln \mathcal{L}
= \vec{C}^T(\alpha, \gamma) \mathbf{C}(\alpha,\gamma)^{-1} \vec{C}(\alpha, \gamma) + \ln\left|\mathbf{C}(\alpha,\gamma)\right|
,
\end{equation}
where $\vec{C}(\alpha, \gamma)$ is a one dimensional array of
$C_\ell^{EB,\mathrm{o} } - \frac{\tan(4\alpha + 4\gamma) }{ 2 }  \left( C_\ell^{EE,\mathrm{o} } - C_\ell^{ BB,\mathrm{o} } \right)
-\frac{\sin(-4\gamma)}{2\cos(4\alpha+4\gamma)}
\left(C_{\ell}^{EE,\mathrm{CMB,th} }b_\ell^2 - C_{\ell}^{BB,\mathrm{CMB,th} }b_\ell^2 
\right)
$ from $\ell_\mathrm{min}$ to $\ell_\mathrm{max}$.

To test our idea, we prepare a dust foreground which follows Eq.~(\ref{eq:EBtoEEBB}) and Eq.~(\ref{eq:BBpropEE}) and ignore synchrotron foreground.
As for the input dust foreground power spectrum, 
we use $C_\ell^{EE,fg}$ calculated from a full-sky map of PySM.
Then we set $C_\ell^{BB,fg} = \xi C_\ell^{EE,fg}$ and  $C_\ell^{EB,fg} = f_c \sqrt{ C_\ell^{EE,fg} C_\ell^{BB,fg}}$.
Using these power spectra, 
we realize a dust foreground map assuming Gaussian fluctuations.
As for the input parameters for the polarized dust emission,
we set $\xi = 0.5$ as suggested in Ref.~\cite{planckdust:2018}, 
and $f_c = 0.01$ which is consistent with observed data~\cite{Abitbol:2015epq}.
From analytical calculation, we can interpret the values of foreground parameters as $\gamma =0.41\,\deg $.

CMB and noise maps are generated in the same way as in Sect.~\ref{sec:SkySim}.
With these inputs,
we determine $\alpha$ and $\gamma$ simultaneously, using the log-likelihood function of Eq.~(\ref{eq:LikelihoodAlphaGammaNonDiagonal});
We use the same binning criteria as in Sect.~\ref{sec:Results}.

We show the recovered $\alpha$, $\gamma$, and their $1\sigma$ uncertainties in Table~\ref{tab:alphagamma}.
As in Sect.~\ref{sec:Results},
we show results of both a single realization and MC simulations with 100 realizations.

We find that input values of $\alpha$ and $\gamma$ are correctly recovered with a single realization
and the estimators are unbiased.
We also find that covariance matrix in Eq.~(\ref{eq:LikelihoodAlphaGammaNonDiagonal}) is valid
because the uncertainties estimated by a single realization
are consistent with the uncertainties derived from  MC simulations.

Because of the simultaneous determination of two parameters,
the uncertainties become larger compared to $\alpha$ only estimation (Table~\ref{tab:alphaCov}).

\begin{table}
	\centering
	\caption{
		Recovered $\alpha $ and $\gamma$ and their $1\sigma$ uncertainty against $\alpha_\mathrm{in}=0.33\,\deg$ and $\gamma_\mathrm{in} = 0.41~(f_c = 0.01\ \mathrm{ and }\ \xi = 0.5)$
		from Eq.~(\ref{eq:LikelihoodAlphaGammaNonDiagonal}).
		We show results of both a single realization and MC simulations.
	}
	\label{tab:alphagamma}
	\begin{tabular}{c c c c}
		\toprule
		\multicolumn{2}{c}{Single realization} & 		\multicolumn{2}{c}{ MC simulations} \\
		\cmidrule(lr){1-2} \cmidrule(lr){3-4}
		$\alpha$ (deg) & $\gamma$ (deg) & $\alpha$ (deg) & $\gamma$ (deg)\\
		\cmidrule(lr){1-2} \cmidrule(lr){3-4}
		$0.33 \pm 0.24$ & $0.46\pm 0.35$& $0.32 \pm 0.25$ & $0.39 \pm 0.33$\\
		\bottomrule
	\end{tabular}
\end{table}

\section{Discussion and conclusion}\label{sec:Conclusion}
In this paper,
we have studied a strategy to determine miscalibrated polarization angles of CMB experiments using the observed $EB$ power spectra of CMB and Galactic foreground emission.
We have extended the methodology of Ref.~\cite{Minami:2019ruj} developed for full-sky observations
to partial-sky observations.
We have corrected the $E$-to-$B$ leakage and $\ell$-to-$\ell$ bin covariance due to partial sky coverage using the framework of NaMaster~\cite{Alonso:2018jzx}.

Applying our method to simulated maps of CMB, realistic foreground emission~\cite{Thorne:2016ifb}, and instrumental noise
with beam smearing similar to the SO LAT~\cite{Ade:2018sbj},
we have found that 
the method correctly recovers the input values of $\alpha$ and estimates its reasonable uncertainty.

We have also developed a method to estimate the intrinsic $EB$ power spectrum of polarized dust emission.
This is possible when $\langle C_\ell^{EB,\mathrm{fg}}\rangle$ is proportional to $\sqrt{\langle C_\ell^{EE,\mathrm{fg}} \rangle \langle C_\ell^{BB,\mathrm{fg}}\rangle}$ and $\langle C_\ell^{BB, \mathrm{fg}} \rangle$ is proportional to $\langle C_\ell^{EE, \mathrm{fg}} \rangle$.
Then we can interpret the $EB$ power spectrum of dust emission as if it were generated by a polarization rotation angle, $\gamma$.
We can determine $\alpha$ and $\gamma$ simultaneously,
as $\alpha$ affects CMB, while $\alpha+\gamma$ affects polarized dust emission. 
We have found that this method correctly recovers the input values of $\alpha$ and $\gamma$ and estimates their reasonable uncertainties.

Though we have assumed homogeneous foreground $EB$ cross correlation parameters, $f_c$ and $\xi$,
we can simply extend our method to deal with spatially varying foreground by replacing them with multipole-dependent ones, i.e., $f_c(\ell)$ and $\xi(\ell)$.
These can be interpreted as $\gamma(\ell)$ and simultaneously determined with $\alpha$.

While we have applied our method to the SO,
we can apply the method to other ground-based observations with a partial sky coverage.
This new framework allows us to determine the miscalibration angle in foreground-dominant frequency bands,
and to detect a non-zero $EB$ power spectrum of polarized dust emission.

We have ignored some detailed experimental parameters.
One of them is 1/f noise which contaminates low multipoles.
Because the miscalibration angles are mainly determined by the information at high multipoles
where cosmic variance is small, 
the effect of 1/f would be small.
We leave studies with more detailed experimental parameters to future work.

\section*{Acknowledgment}
We thank E.~Komatsu and Y.~Chinone for useful discussion and feedback on this project,
and S.~Takakura, M.~Murata, A.~Kusaka, M.~Hasegawa, and O.~Tajima for comments on the draft.
We thank N.~Krachmalnicoff for helping the use of PySM.
We acknowledge the use of NaMaster and thank D.~Alonso and M.~R.~Becker for their help to use the code.
This work was supported in part by the JSPS Core-to-Core Program, A. Advanced Research Networks and my wife.
\appendix

\section{Validation of the covariance matrix of NaMaster}
\label{sec:ValidCov}
In this section,
we show two tests to validate the covariance matrix calculated by NaMaster~\cite{Alonso:2018jzx}:
(1) Validation of diagonal elements;
(2) Validation of off-diagonal elements with binning.

We validate the covariance matrix calculated analytically with the NaMaster framework  by comparing with that from Monte-Carlo (MC) simulations.
Because it needs enormous data volume to simulate off-diagonal covariance matrix elements with MC simulation,
we validate diagonal elements of the covariance matrix with $N= 100$ samples of \textit{unbinned} power spectrum,
and validate off-diagonal ones with $N=1000$ MC samples of \textit{binned} power spectrum. 

In the MC simulations,
we generated $N$ full-sky maps assuming Gaussian distributions with zero mean as we have described in Sect.~\ref{sec:SkySim}.

\subsection{Validation of diagonal elements}\label{sec:ValidDiagonal}
We compare diagonal elements of the covariance matrix $\mathbf{C}(\alpha)$ (\ref{eq:Covariance}) calculated analytically with the NaMaster framework 
and those calculated with $N = 100$ MC-simulated samples.

\begin{figure}
	\centering
	\includegraphics[width=\linewidth]{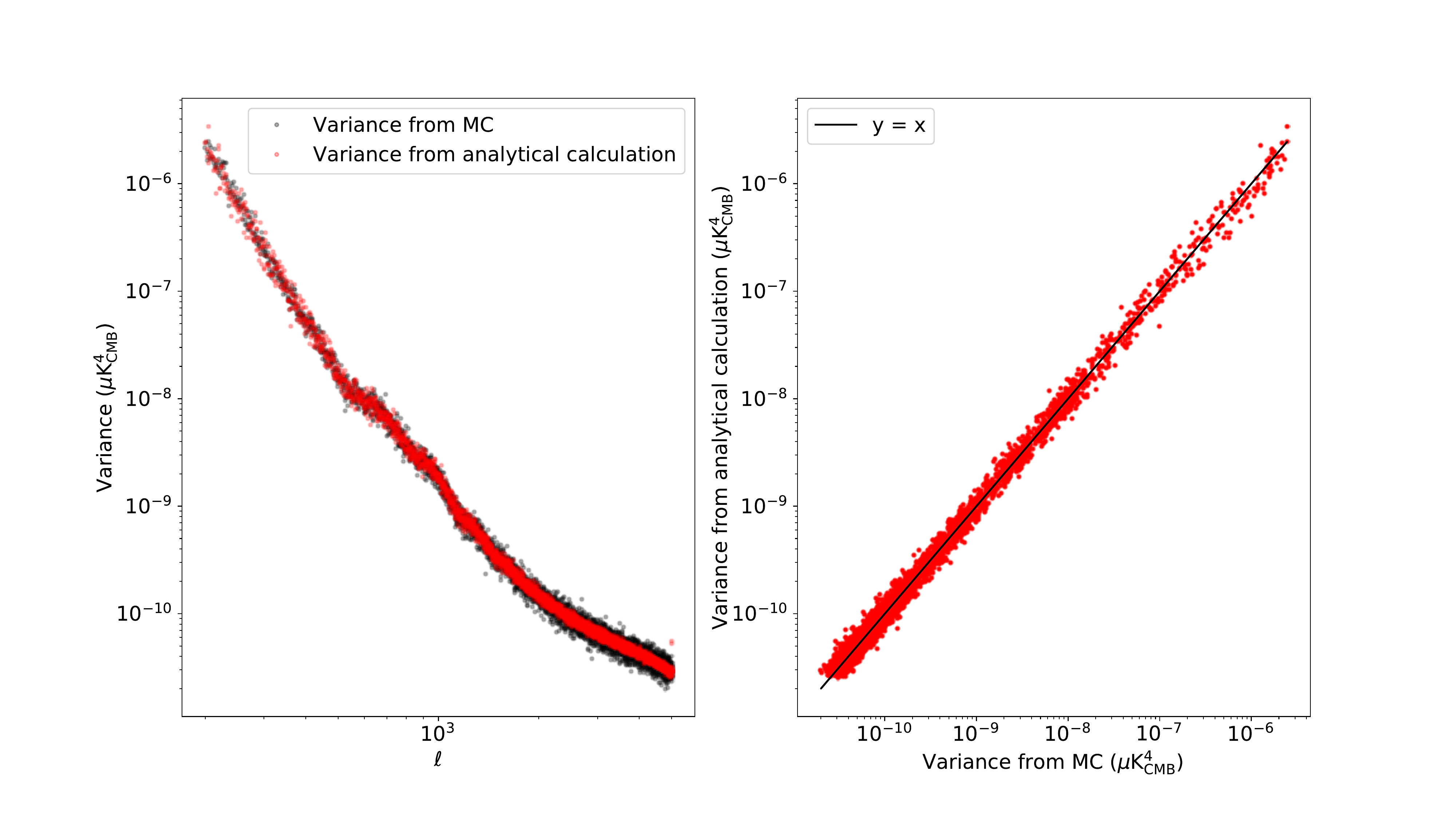}
	\caption{
		Left: diagonal elements of the covariance matrix derived from $N=100$ MC-simulated samples (black dots) and those analytically calculated by NaMaster (red dots).
		Right: $\ell$-by-$\ell$ comparison of the diagonal elements of the covariance matrices.
		For each $\ell$-bin, the diagonal element analytically calculated by NaMaster is plotted against that derived from the MC-simulated samples (red dots).
		The black solid line shows the case that the all diagonal elements derived from two methods agree exactly.
	}
	\label{fig:Covariance}
\end{figure}

We show the comparison in Figure~\ref{fig:Covariance}.
We find that analytically-calculated covariance-matrix elements are consistent with MC-simulated covariance-matrix elements at all multipoles.

\subsection{Validation of off-diagonal elements}
\label{sec:ValidNonDiagonal}
We compare off-diagonal elements of the covariance matrix $\mathbf{C}(\alpha)$ (\ref{eq:Covariance}) 
calculated analytically with the NaMaster framework with
those from $1000$ MC-simulated samples.
In the calculation of both analytical covariance-matrix and MC-simulated power spectra, 
the covariance is binned with a bin size of $\Delta\ell=8$ with $\ell$ range of $(\ell_\mathrm{min}, \ell_\mathrm{max}) = (200, 5000)$.

\begin{figure}
	\centering
	\includegraphics[width=\linewidth]{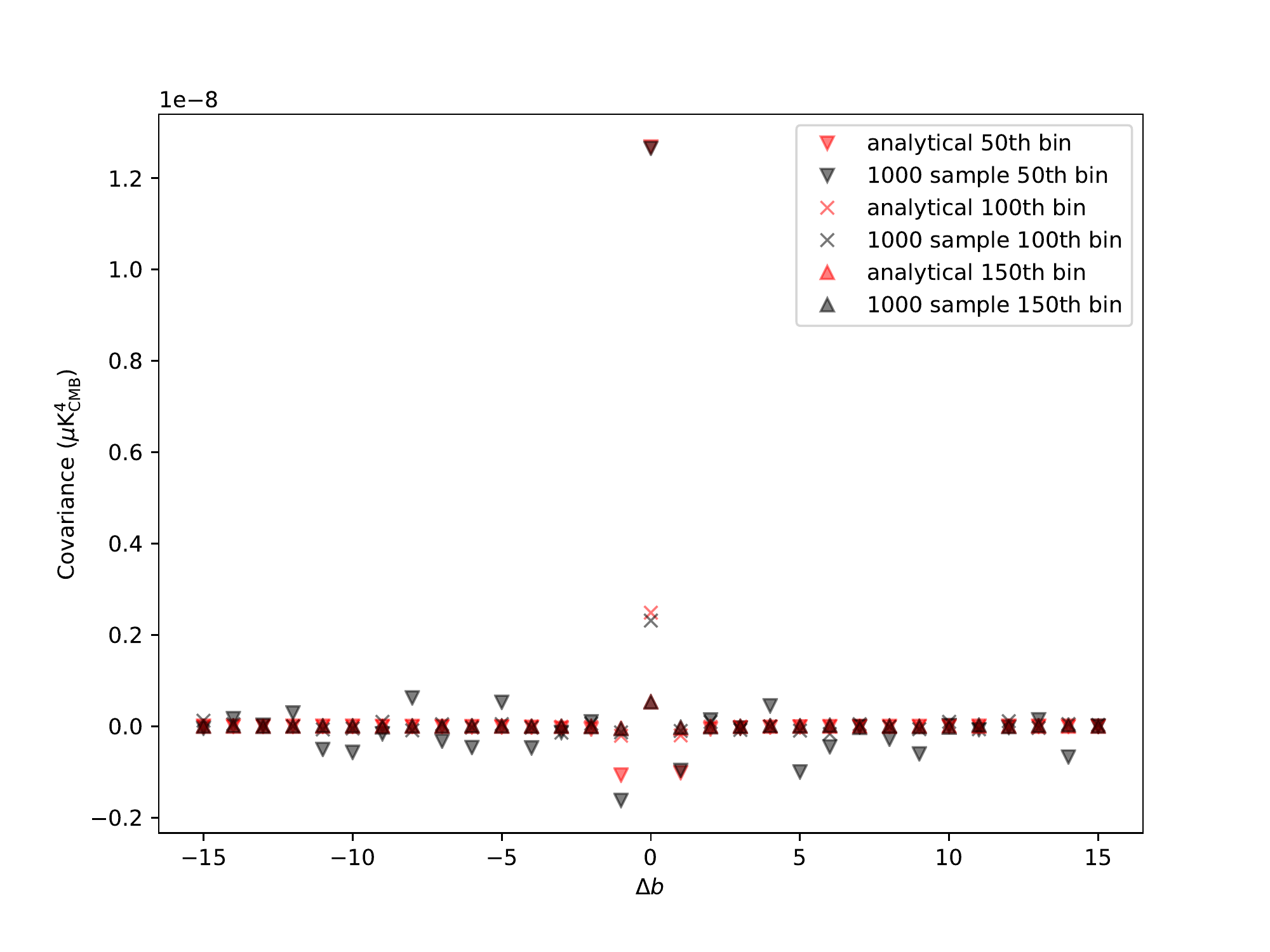}
	\caption{
		The covariance-matrix elements against the distance between bins, $\Delta b$.
		The black markers are derived from 1,000 MC simulations, 
		while the red markers are analytically calculated by NaMaster.
	}
	\label{fig:Covariance_Nondiag}
\end{figure}

We show the comparison of the off-diagonal elements of covariance matrices in Figure~\ref{fig:Covariance_Nondiag}.
We find that
analytically-calculated covariance-matrix elements are
consistent with MC-simulated covariance-matrix elements
at all distances between bins, $\Delta b$.
And, we can see that $|\Delta b| \sim 1 $ has a negative covariance.
Therefore,
ignoring the off-diagonal covariance-matrix elements (as in Eq.~(\ref{eq:LikelihoodAlphaDiagonal}) and Eq.~(\ref{eq:LikelihoodTraditional})) overestimates uncertainties on the angles.

\bibliographystyle{ptephy}
\bibliography{references}

\begin{thebibliography}{10}

\bibitem{Minami:2019ruj}
Yuto Minami, Hiroki Ochi, Kiyotomo Ichiki, Nobuhiko Katayama, Eiichiro Komatsu,
  and Tomotake Matsumura, PTEP, {\bf 2019}(8), 083E02 (2019),
  {{arXiv:1904.12440}}.

\bibitem{Kamionkowski:2015yta}
M.~Kamionkowski and E.~D. Kovetz, Ann. Rev. Astron. Astrophys., {\bf 54},
  227--269 (2016),  {{arXiv:1510.06042}}.

\bibitem{Zaldarriaga:1998ar}
Matias Zaldarriaga and Uros Seljak, Phys. Rev., {\bf D58}, 023003 (1998),
  {{arXiv:astro-ph/9803150}}.

\bibitem{Hu:1999vq}
Wayne Hu, Astrophys. J., {\bf 529}, 12 (2000),  {{arXiv:astro-ph/9907103}}.

\bibitem{Dvorkin:2009ah}
Cora Dvorkin, Wayne Hu, and Kendrick~M. Smith, Phys. Rev., {\bf D79}, 107302
  (2009),  {{arXiv:0902.4413}}.

\bibitem{Bunn:2002df}
Emory~F. Bunn, Matias Zaldarriaga, Max Tegmark, and Angelica de~Oliveira-Costa,
  Phys. Rev., {\bf D67}, 023501 (2003),  {{arXiv:astro-ph/0207338}}.

\bibitem{Smith:2006vq}
Kendrick~M. Smith and Matias Zaldarriaga, Phys. Rev., {\bf D76}, 043001 (2007),
   {{arXiv:astro-ph/0610059}}.

\bibitem{Komatsu:2010fb}
E.~Komatsu et~al., Astrophys. J. Suppl., {\bf 192}, 18 (2011),
  {{arXiv:1001.4538}}.

\bibitem{Keating:2012ge}
B.~Keating, M.~Shimon, and A.~Yadav, Astrophys. J., {\bf 762}, L23 (2012),
  {{arXiv:1211.5734}}.

\bibitem{Hazumi2019}
M.~Hazumi et~al., Journal of Low Temperature Physics, {\bf 194}(5), 443--452
  (Mar 2019).

\bibitem{Ade:2018sbj}
Peter Ade et~al., JCAP, {\bf 1902}, 056 (2019),  {{arXiv:1808.07445}}.

\bibitem{Ben:2018}
Benjamin Westbrook et~al., Journal of Low Temperature Physics, {\bf 193} (09
  2018).

\bibitem{Hui:2018cvg}
Howard Hui et~al., Proc. SPIE Int. Soc. Opt. Eng., {\bf 10708}, 1070807 (2018),
   {{arXiv:1808.00568}}.

\bibitem{abazajian2019cmbs4}
Kevork Abazajian et~al.,
\newblock Cmb-s4 decadal survey apc white paper (2019),  {{arXiv:1908.01062}}.

\bibitem{Alonso:2018jzx}
David Alonso, Javier Sanchez, and Anže Slosar, Mon. Not. Roy. Astron. Soc.,
  {\bf 484}(3), 4127--4151 (2019),  {{arXiv:1809.09603}}.

\bibitem{Carroll:1998zi}
S.~M. Carroll, Phys. Rev. Lett., {\bf 81}, 3067--3070 (1998),
  {{arXiv:astro-ph/9806099}}.

\bibitem{Lue:1998mq}
A.~Lue, L.-M. Wang, and M.~Kamionkowski, Phys. Rev. Lett., {\bf 83}, 1506--1509
  (1999),  {{arXiv:astro-ph/9812088}}.

\bibitem{Feng:2004mq}
B.~Feng, H.~Li, M.~Li, and X.~Zhang, Phys. Lett., {\bf B620}, 27--32 (2005),
  {{arXiv:hep-ph/0406269}}.

\bibitem{Feng:2006dp}
B.~Feng, M.~Li, J.-Q. Xia, X.~Chen, and X.~Zhang, Phys. Rev. Lett., {\bf 96},
  221302 (2006),  {{arXiv:astro-ph/0601095}}.

\bibitem{Liu:2006uh}
G.-C. Liu, S.~Lee, and K.-W. Ng, Phys. Rev. Lett., {\bf 97}, 161303 (2006),
  {{arXiv:astro-ph/0606248}}.

\bibitem{planckdust:2016}
Planck Collaboration~Int. XXX, Astron. Astrophys., {\bf 586}, A133 (2016),
  {{arXiv:1409.5738}}.

\bibitem{planckdust:2018}
Planck~Collaboration XI (2018),  {{arXiv:1801.04945}}.

\bibitem{Aghanim:2016fhp}
Planck Collaboration~Int. XLIX, Astron. Astrophys., {\bf 596}, A110 (2016),
  {{arXiv:1605.08633}}.

\bibitem{Hivon:2002}
Eric Hivon, Krzysztof~M. Gorski, C.~Barth Netterfield, Brendan~P. Crill, Simon
  Prunet, and Frode Hansen, The Astrophysical Journal, {\bf 567}(1), 2--17 (mar
  2002).

\bibitem{Elsner:2016}
Franz Elsner, Boris Leistedt, and Hiranya~V. Peiris, Monthly Notices of the
  Royal Astronomical Society, {\bf 465}(2), 1847--1855 (10 2016),
  {{http://oup.prod.sis.lan/mnras/article-pdf/465/2/1847/8364799/stw2752.pdf}}.

\bibitem{Efstathiou:2004}
G.~Efstathiou, Monthly Notices of the Royal Astronomical Society, {\bf 349}(2),
  603--626 (04 2004),
  {{http://oup.prod.sis.lan/mnras/article-pdf/349/2/603/2826253/349-2-603.pdf}}.

\bibitem{Brown:2004jn}
Michael~L. Brown, P.~G. Castro, and A.~N. Taylor, Mon. Not. Roy. Astron. Soc.,
  {\bf 360}, 1262--1280 (2005),  {{arXiv:astro-ph/0410394}}.

\bibitem{Thorne:2016ifb}
B.~Thorne, J.~Dunkley, D.~Alonso, and S.~Naess, Mon. Not. Roy. Astron. Soc.,
  {\bf 469}(3), 2821--2833 (2017),  {{arXiv:1608.02841}}.

\bibitem{Katayama:2011eh}
N.~Katayama and E.~Komatsu, Astrophys. J., {\bf 737}, 78 (2011),
  {{arXiv:1101.5210}}.

\bibitem{Lewis:2000}
A.~{Lewis}, A.~{Challinor}, and A.~{Lasenby}, Astrophys. J., {\bf 538},
  473--476 (August 2000),  {{astro-ph/9911177}}.

\bibitem{Aghanim:2018eyx}
N.~Aghanim et~al. (2018),  {{arXiv:1807.06209}}.

\bibitem{Aumont:2018epb}
J.~Aumont, J.~F. Macías-Pérez, A.~Ritacco, N.~Ponthieu, and A.~Mangilli
  (2018),  {{arXiv:1805.10475}}.

\bibitem{Abitbol:2015epq}
M.~H. Abitbol, J.~C. Hill, and B.~R. Johnson, Mon. Not. Roy. Astron. Soc., {\bf
  457}(2), 1796--1803 (2016),  {{arXiv:1512.06834}}.

\end{thebibliography}

\end{document}